**Forming Weakly Interacting Multi Layers of Graphene by using Atomic Force Microscope Tip Scanning and Evidence of Competition Between Inner and Outer Raman Scattering Processes Piloted by Structural Defects**

C. Pardanaud[*1], A. Merlen[2], K. Gratzer[3], O. Chuzel[3], D. Nikolaievskyi[1, 3], L. Patrone[2], S. Clair[2], R. Ramirez Jimenez[4,5], A. de Andrés[5], P. Roubin[*1], J.-L. Parrain[3]

[1]*Aix Marseille Université, CNRS, PIIM UMR 7345, 13397, Marseille, France*
[2]*IM2NP, UMR CNRS 7334, Aix-Marseille Université, Université de Toulon, France*
[3]*Aix Marseille Univ, CNRS, Centrale Marseille, iSm2, Marseille, France*
[4]*Departamento de Física, Escuela Politecnica Superior, Universidad Carlos III de Madrid, Avenida Universidad 30, Leganes, 28911 Madrid, Spain*
[5] *Instituto de Ciencia de Materiales de Madrid. Consejo Superior de Investigaciones Científicas. Cantoblanco 28049 Madrid, Spain*

Corresponding author: cedric.pardanaud@univ-amu.fr, pascale.roubin@univ-amu.fr

**Abstract**

We report on an alternative route based on nanomechanical folding induced by AFM tip to obtain weakly interacting multi-layer graphene (wi-MLG) from a chemical vapor deposition (CVD) grown single-layer graphene (SLG). The tip first cuts, then pushes and folds graphene during zig-zag movements. The pushed graphene has been analyzed using various Raman microscopy plots – $A_D/A_G \times E_L^4$ vs $\Gamma_G$, $\omega_{2D}$ vs $\Gamma_{2D}$, $\Gamma_{2D}$ vs $\Gamma_G$, $\omega_{2D+/-}$ vs $\Gamma_{2D+/-}$, and $A_{2D-}/A_{2D+}$ vs $A_{2D}/A_G$. We show that the SLG in plane properties are maintained under the folding process and that a few tens of graphene layers are stacked, with a limited amount of structural defects. A blue shift of about 20 cm$^{-1}$ of the 2D band is observed. The relative intensity of the 2D- and 2D$_+$ bands have been related to structural defects, giving evidence of their role in the inner and outer processes at play close to the Dirac cone.

**Graphical abstract:**

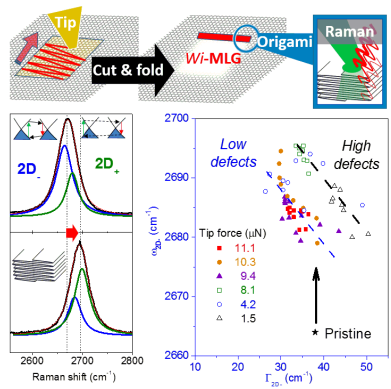

Since its first fabrication in 2004,[1] single layer graphene (SLG) has been widely studied. Many efforts have been done to synthetize it,[2] developing characterization techniques to understand how atomic structure and defects affect phonons and electrons which themselves drive heat and electronic transport properties.[3] The main effects arise from the Dirac points at the K point in the Brillouin zone (BZ), where the π and π* bands cross linearly at the Fermi level, forming the so-called Dirac cones that lead the electrons to behave as massless particles with a Fermi velocity reported as high as ≈ 300 times lower than the speed of light in vacuum. Changing in a controlled way the behavior of these electrons (by band gap opening, Fermi velocity tuning,[4, 5] etc.) is a major concern for the future of nanoelectronics,[6] strain engineering[7] and sensing applications[8].

Graphenic family members also inherit some of the astounding properties of graphene planes, such as graphite, amorphous carbon, nanotubes, nanocones, nanoribbons, graphene oxide, etc. Among this family, three-dimensional porous networks of planar graphene[9] have been recognized to play an important role in the forthcoming technologies.[10, 11] It has been shown recently that topology, curvature and pore properties lead to changes in the electronic and vibrational structures[12] or that introducing distortions by hole doping influences the density of states at the Fermi level. Stable Z-shape folded graphene produced using AFM tip has been obtained very recently[13, 14] opening the area of kirigami/origami of graphene.[15] Tailoring the 3-dimension architecture of graphene is a key point in controlling some of its properties[16] and the formation of multi-layer graphene (MLG) is of importance for this purpose. It has been shown that the number of layers and the way they are stacked deeply modify the nature of the charge carrier and, therefore, the electronic properties.[17] MLG have already been used for many applications, for example, added to metal oxides in highly sensitive sensors allows diminishing the working temperature.[18] Due to the ability of controlling its lateral size during synthesis, MLG is also a material of choice for interacting with biological materials.[19] Several production methods have been used in the past, as classical mechanical cleavage, chemical vapor deposition (CVD) techniques,[20] thermal treatment of multiwalled carbon nanotubes[21] or layer by layer growth.[22]

In this letter, we report on an alternative route to obtain MLG using nanomechanical folding induced by AFM tip. The movement of the probe tip is controlled in order to initially cut a SLG and then fold it to form MLG. Interestingly, the folding is created at the heart of the layer and independently from the edges. The interlayer interaction in this folded graphene is expected to differ substantially from that derived from the previously mentioned fabrication methods for MLG. Indeed, Raman microscopy analysis, performed to characterize the properties of these MLG, evidence that this folded graphene behaves as weakly interacting MLG (wi-MLG), without doping or strain, and emphasize the role of the structural defects caused by the folding process.

**Raman spectroscopy background**

Mainly due to the double resonance (DR) mechanism, which is based on electron scattering by incoming light that selects phonons that satisfy wavevector and energy conservation close to K and K' points of the BZ, resonance Raman microscopy of graphenic materials is a key characterization technique[23]. It can give information on structural, mechanical and electronic properties. A typical Raman spectrum of graphene is dominated by two or three intense bands plus less intense combination bands. The in-plane vibration G band, with $E_{2g}$ symmetry at the $\Gamma$ point of the BZ, lies at 1582 cm$^{-1}$; the defect induced D band, with $A_{1g}$ symmetry, lies close to 1350 cm$^{-1}$ using a laser wavelength $\lambda_L$=514 nm and the two-phonon double resonant 2D band, lies close to 2680 cm$^{-1}$ with $\lambda_L$=514 nm,[24] both close to the K points. Whereas the D band is activated by defects in the crystal structure, the 2D band, involving the scattering by two phonons, is always present in the spectrum.[25] Due to both the DR mechanism and the phonon dispersion, the D and 2D band positions depend on the laser wavelength used.[26] The 2D band shape is found Lorentzian for supported graphene but is asymmetric for free standing graphene, being fitted with two Lorentzian curves[27] with full widths in the range of 25 to 50 cm$^{-1}$.[28] This asymmetric profile is explained in the framework of the DR mechanism: the 2D band involves phonons with wavevectors that are close to the KK' wavevector value. It means that phonons which contribute to the Raman cross section via the DR mechanism can have a wavevector either lower or higher than KK', leading to inner or outer scattering processes, respectively.[25, 26, 29-31] As both electron and phonon dispersions are not symmetric close to this high symmetry BZ line, the DR mechanism selects phonons that have slightly different energies, leading to the two contributions forming the asymmetric profile. The subband at the lowest wavenumber (2D-) is attributed to the inner process whereas the band at the highest wavenumber (2D+) is attributed to the outer process. A level of doping higher than $2\ 10^{11}$ cm$^{-2}$, due to interaction with the substrate or using an electrostatic field, destroys this bimodal shape.[30] A lower value of doping is found to modify the distance between the 2D$_-$ and 2D$_+$ bands.[31] The intensity ratio of these bands has been found to be close to $\approx 3.5$ with $\lambda_L$=514 or 633 nm, and the frequency difference in the range 6.6-12 cm$^{-1}$.[30, 32] Higher doping can affect also other parts of the Raman spectrum: position and width of the G band ($\omega_G$ and $\Gamma_G$, respectively), and relative intensity ratio $I_{2D}/I_G$.[33] Mechanical strain can also affect the Raman spectrum.[34] The 2D band position plotted as a function of the G band position (labelled as $\omega_{2D}$ vs $\omega_G$ plot) has been found to be able to disentangle electron or hole doping from macroscopic strain effects (for doping higher than $\approx 10^{12}$ cm$^{-2}$).[35-37] A slope close to 2 in that plot reveals pure macroscopic strain effect, a slope close to 0.7 reveals pure doping and intermediate values reveal a combination of the two effects that can be disentangled using basic algebra.

Depending on the density of in plane defects or on the doping level, band widths and intensity ratios of D and 2D bands over G band are also used to better characterize material properties [38]. For example, in the case of graphite, $I_D/I_G$ has been frequently used to quantify the aromaticity, i.e. the in plane crystalline quality: $I_D/I_G$ the higher, the aromatic domain size ($L_a$) the smaller.[39] The shape of the 2D band of graphitic materials contains information on the stacking, as reported early.[40] For bilayer graphene (BLG), the shape of the 2D band (with a width close to 50 cm$^{-1}$ or higher) is more complex: it is split in four subbands which are related to scattering processes involving also interaction of the $\pi$ and $\pi$* orbitals of the two sublayers.[41] Introducing a stacking default in BLG can change drastically its properties as well as the 2D band shape.[42] Concerning the G band intensity, an increase up to ×60 very sensitive to both $\lambda_L$ and the twisting angle is reported to be due to sublayer interaction.[43] For MLG, the 2D band, giving access to electronic structure information through the DR mechanism, allows to retrieve information from the way graphene sheets are connected. The 2D band shape changes from a quasi-Lorentzian or bi-Lorentzian shape for a SLG, to a composition of several bands for MLG (at least four bands for BLG).[44] However, for not well-stacked MLG, it has been shown that the profile is close to that of a SLG since stacking a SLG on a SLG does not necessarily leads to the properties of a BLG. Moreover, a folded SLG on top of a SLG leads to two separated SLG, but with a lower Fermi velocity.[4] Compared to SLG, the 2D band blueshifts (in the range 4 - 12 cm$^{-1}$), whereas the G band shifts by less than 1 cm$^{-1}$. Similar results have been found for misoriented MLG (up to 6 layers) that lead to Lorentzian profiles and blue shifts of the 2D band.[28] Finally, we mention the weak asymmetric combination band D + D'' at 2450 cm$^{-1}$, originated by the inner scattering process plus phonons coming from the ΓK high symmetry line of the BZ,[45] whose shape and position depend on the number of stacked layers.

**Experimental procedures and Raman analysis**

SLG samples have been obtained from CVD graphene deposited onto Cu and transferred by means of PMMA spin coating on a Si wafer (with a Si native oxide layer of a few nm). Acetone was used to remove the PMMA. In order to obtain the best SLG, labelled as pristine hereafter, we tested different experimental conditions varying $CH_4$ and $H_2$ flows, the baking time during spin coating, the nature of the etchant and the time acetone was applied to remove PMMA (See supplementary information for more details). Measurements were performed in air. Raman spectra were obtained using a Horiba Jobin Yvon HR800 set up with an excitation wavelength $\lambda_L$=514 nm, a ×100 objective (numerical aperture of 0.9, i.e. a theoretical spot radius of 0.34 μm), a 600 grooves/mm grating and 5 mW power. Resolution is about 1 cm$^{-1}$. We selected a sample with the following Raman features: $I_{2D}/I_G$ = 2.3, $\omega_{2D}$ = 2678 cm$^{-1}$, $\omega_G$ = 1581 cm$^{-1}$, $\Gamma_{2D}$ = 37 cm$^{-1}$ (falling in the range admitted for SLG 2D band[28]) and $\Gamma_G$ = 25 cm$^{-1}$. Raman maps were performed with a lateral displacement step of 250 nm and data were extracted from these maps. In specific cases, to increase the signal to noise ratio, spectra were averaged from several relevant μm$^2$ zones, after checking it was homogeneous.

To cut and modify the pristine SLG, we used an atomic force microscope (AFM) tip, as introduced in [46, 47]. Starting at the bottom left of a 4x4 μm$^2$ square and finishing at the top right, the tip scans the square with zig zag movements in 512 lines of 4 μm (with an overlapping of the contact zone from one passage of the tip to the other, as the tip radius is ≈ 10 nm) at a speed of 8 μm s$^{-1}$. The vertical force applied was set constant for a given square and several forces, ranging from 0.2 to 11.1 µN, were tested on 25 separate squares. For a square, the first passage is supposed to cut the SLG on a 4 µm length whereas the subsequent zig-zag movements have two roles. First, they constantly apply a strain in the horizontal plane, pushing and folding graphene. Second, they help in cutting graphene from the edges perpendicular to the direction of the first passage. The AFM used in a tapping mode is an Agilent 5500 microscope, used in liquid medium (acetonitrile + hydrogen peroxide). The tip is a commercial silicon tip (AppNano ACT type, cantilever length 125 µm, nominal frequency 300 kHz) coated with the 1,4,7-triazacyclononane ligated manganese complex as explained in.[46, 47] Note that investigating the role of the coating, and its related chemical influence is beyond the scope of this letter.

**First Raman analysis: evidence for the formation of a weakly interacting MLG**

Graphene is cut for forces equal or higher than 1.5 µN. For these forces, graphene is pushed on the top and on the sides of the scanned square, as can be seen in the image obtained by Atomic Force Microscope (AFM), Figure 1a. Graphene pushed on top appears as a rectangular band with dimensions of ~ 300 to 600 nm wide and 4 µm long, as measured from raw images, covering a projected surface area of $2.4 \pm 0.2$ µm$^2$. Graphene pushed on sides appears as small isolated patches. Figure 1b displays a typical line profile showing that the apparent height is about 30 to 45 nm in average, with peaks locally reaching up to 75 nm high. Figure 1c displays a typical Raman spectrum of pushed graphene in comparison with a typical one of pristine graphene (outside the scanned zones). The main bands of the pushed graphene spectrum are the G and 2D bands, with a G band width similar to that of the pristine SLG, showing that the main graphene properties remain during the process. A small D band appears, which shows that some defects are introduced (as discussed below). The 2D band position is shifted from 2678 cm$^{-1}$ for the pristine graphene to $\approx$ 2700 cm$^{-1}$ for the pushed one, with a shape which is a bimodal Lorentzian profile. Figure 1d shows a full map of the studied graphene area, the scanned zones being indicated by the white dotted squares. The first, second and third squares are not commented in this study as the tip force was not enough to remove graphene. In this map, the relative intensity of the G band is displayed for each force (from 0.2 to 11.1 µN) in a logarithmic scale. The surrounding areas (in light blue) have a relative intensity of $I_G/I_{Gpristine} = 1$ (arbitrary units, 0 in log scale) and correspond to pristine graphene. The scanned zones (in dark blue) have intensities at the noise level meaning that graphene was completely removed by the tip movement. The top zones (in yellow and red) display the highest G band intensities, with a multiplication factor when compared to pristine graphene up to $I_G/I_{Gpristine} \approx 100$ (arbitrary units, 2 in log scale), while the side zones (in cyan) have a moderate multiplication ($I_G/I_{G\,pristine} \approx 3$).

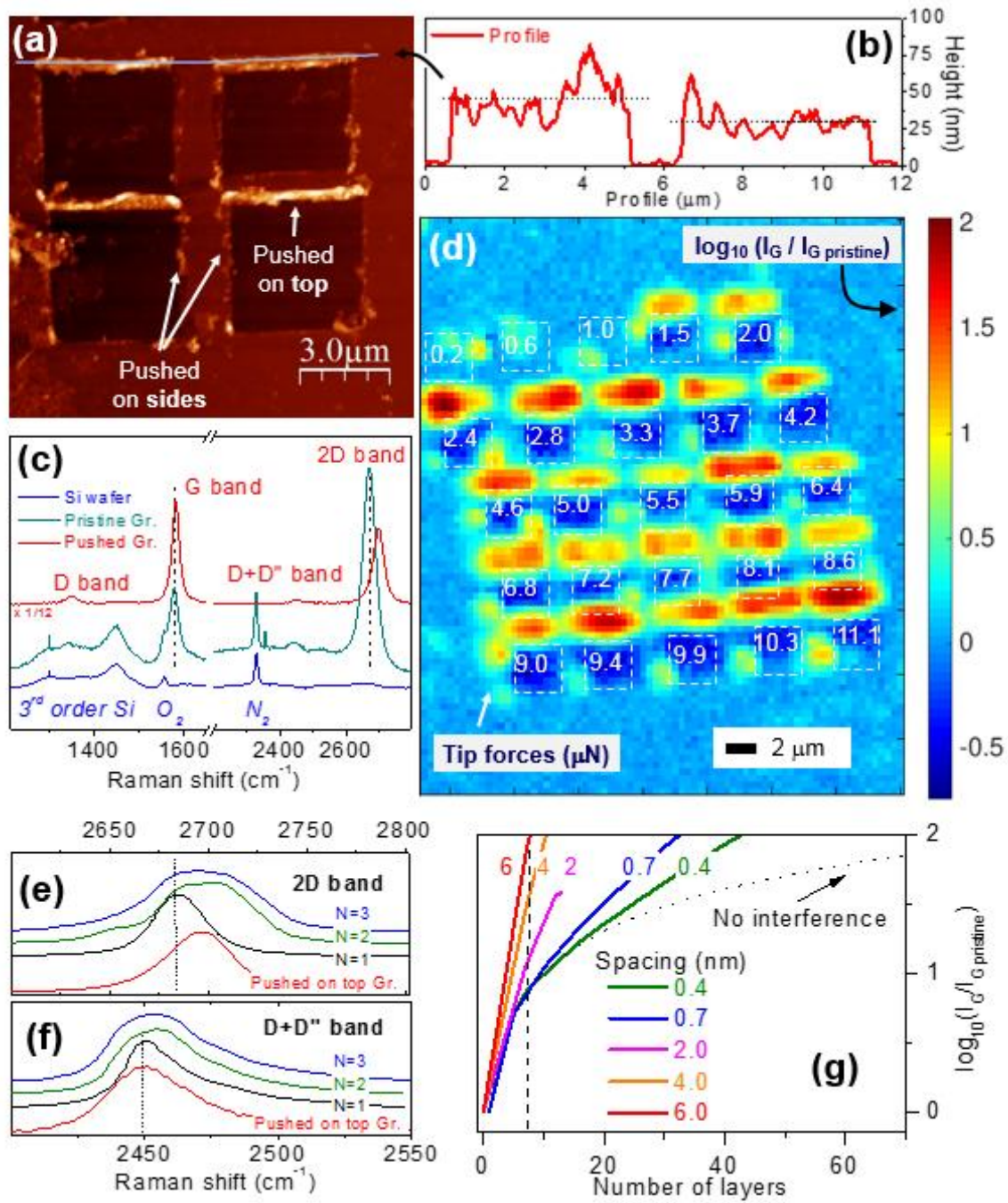

**Figure 1.** Cut and pushed graphene after scanning probe nanolithography. (a, b) AFM image and corresponding height profile along the blue line. (c) Raman spectrum before (green) and after (red) the tip scan (in blue that of the substrate). (d) Raman map of the G band intensity. Numbers in the white squares are the tip forces applied in μN. (e) 2D band of the top pushed graphene compared to N-layer graphene with N=1 (SLG), N=2 (BLG) and N=3 (TLB), adapted from.[44] (f) D + D" band shape of the pushed graphene compared to N=1 (SLG), N=2 (BLG) and N=3 (TLB) graphene, adapted from.[45] We corrected the effect of wavelength excitation from 2.33 eV (in[45]) to 2.41 eV (this work) by using the slope of -32.9 cm$^{-1}$/eV given in.[45] (g) G-band intensity modeled by reflection and diffusion processes in MLG.

The 2D band (Figure 1) behaves neither like bi-, tri-, quadri-layer graphene, nor like graphite[28, 44, 48], but behaves like SLG, with a single or a bimodal Lorentzian shape, which is an indication of a stacking misorientation, and then of a weaker interaction between planes. Compared to SLG, its position is blue shifted by 10 to 20 cm$^{-1}$, may be due to a reduction of the Fermi velocity[4]. Figure 1f displays the weak D + D'' band which also compare better with SLG, although with a larger width, than with bi-, tri- or quadri-layer graphene. Because both the laser spot and the folded zone are hundreds of nanometers large, it could be reasonable to think that all the Raman spectra of folded graphene displayed in this study are a combination of the folded graphene itself and the surrounding SLG. However, this is not the case and one could consider that the Raman spectra presented here are mainly due to folded graphene. There are two reasons: first because the overlapping has been chosen the smallest as possible (spectra and the relevant spectroscopic data have been extracted as far as possible from the edges of the folded zone) and second, because the signal coming from the folded graphene is much more intense (up to two orders of magnitude) than the signal coming from SLG.

The observations reported above suggest that the folded graphene planes are stacked but weakly interacting. To comfort this picture, we performed calculations on the influence of the folding on the G-band intensity. Figure 1g displays the G-band intensity obtained by modeling the interference of multiple internal reflections of the incident and scattered light involved in the Raman process for MLG films on silicon, using a modification of the approach used for SLG on multilayered substrates[49, 50]. We have considered a simple geometry: stacked graphene layers oriented tangentially to the surface, as suggested by the Z folding shape found in[13, 14] that occurs because of van der Waals interaction. If we consider that the full 4×4 μm$^2$ initial graphene square is folded homogeneously onto 2.4 μm$^2$, this would lead to a maximum of 4×4/2.4≈7 layers. With an apparent height of 45 nm (in average) or 75 nm (max value) as measured by AFM, this gives an interlayer spacing of 6.4 nm and 10.7 nm respectively. This is consistent with the calculations shown in Figure 1g and the measured Raman intensity enhancement of ~100 for a 7-multilayer graphene.

A different geometry could be also considered. As we will show below, a coherence length of 50 nm is expected in the folded graphene. If we now consider that the full 4×4 µm$^2$ initial graphene square is folded homogeneously onto a 4 µm × 50 nm area, this would lead to a maximum of 80 layers. With an apparent height of 45 nm (in average) or 75 nm (max value) as measured by AFM, this would give an interlayer spacing of 0.6 nm and 0.9 nm respectively. According to the calculations shown in Figure 1g, such high number of layers would give a too high intensity enhancement. This case requires thus that no interference is considered, i.e. that a non-flat lying geometry is obtained for a 80 multilayer graphene. We have to mention that the structure of the folded graphene is highly heterogeneous, as illustrated by the AFM line profile of Figure 1b and the Raman intensity of Figure 1c. It is thus expected that a combination of different geometries could be found inside the folded graphene band, with different numbers of layers and different interlayer spacing, Figure 1g showing that a large range of geometries could give similar significant intensity enhancements, even if it is very improbable that 80 layers are present. Note that according to the calculations, introducing small random variations of the interlayer distance could also modifies intensities, up to a maximum of ± 30% (not shown). Other factors can also affect the intensity of the G band, that will be discussed in the last part of this paper. In the next part, we will discuss structural, electronic and mechanical properties of this MLG with the help of the relations between the 2D and G band features by plotting $\omega_{2D}$ vs $\omega_G$ and $\Gamma_{2D}$ vs $\Gamma_G$.

To check the presence of doping or strain, we display a $\omega_{2D}$ vs $\omega_G$ plot in Figure 2a where the two straight lines corresponding to the effect of electron and hole doping[51, 52] and to the effect of strain[53] are also plotted. $\omega_{2D}$ vs $\omega_G$ plots are sensitive to doping larger than ≈ 1 10$^{12}$ cm$^{-2}$ and strain larger than ≈ ± 0.1%.

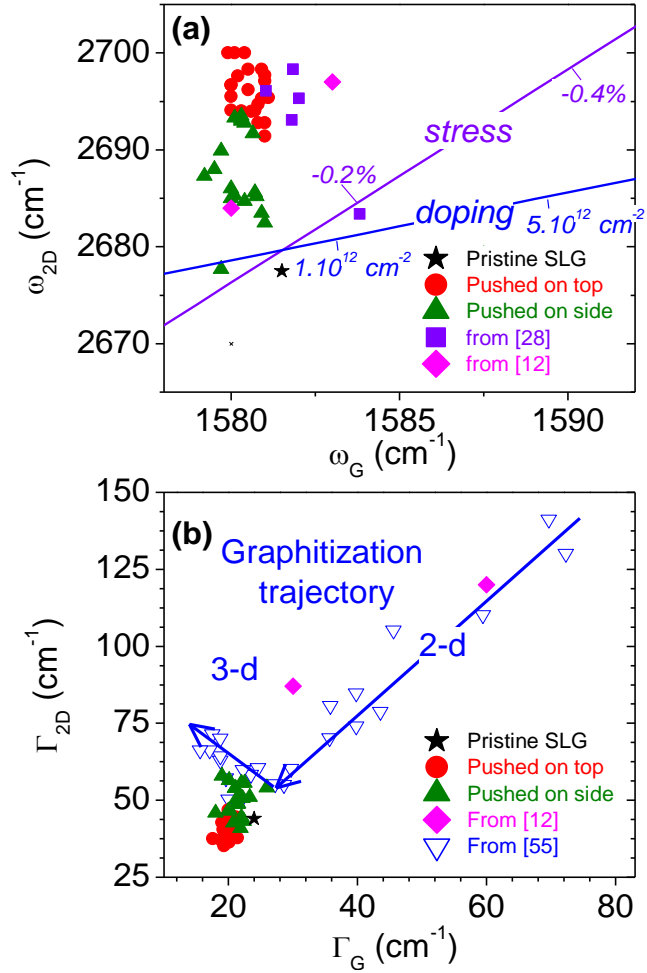

**Figure 2.** (a) Doping and strain characterization with a $\omega_{2D}$ vs $\omega_G$ plot. Data from Bayle *et al*[28] have been $\lambda_L$-corrected using a 2D band dispersion of 100 cm$^{-1}$ / eV. Straight lines correspond to the behavior under doping (blue) or strain (violet).[35-37] (b) Structural characterization with a $\Gamma_{2D}$ vs $\Gamma_G$ plot.

Consistently, the pristine graphene data point is close to the intersection of these two lines which corresponds to suspended graphene (no doping and no strain). The wi-MLG folded graphene data points of this work follow none of these two lines. They draw roughly a vertical line: $\omega_G$ is nearly constant, in the range 1579 - 1581 cm$^{-1}$, while $\omega_{2D}$ is in the range 2677 - 2702 cm$^{-1}$. Values obtained for the side pushed graphene are smaller than for the top pushed graphene, the maximum blue shift compared to pristine graphene being ≈ 25 cm$^{-1}$. Similar blue shifts have been observed previously for not well stacked graphene layers[28] and attributed to a reduction of the Fermi velocity.[4, 54] We then conclude here that pushed graphene has a lower Fermi velocity than pristine graphene (about 0 to 6% less for side pushed graphene and about 5 to 11% less for top pushed graphene) whereas neither doping nor strain effects are observed within Raman sensitivity.

$I_{2D}/I_G$ was found to vary from $\approx$ 0.4 to 3 without distinction between top and side pushed graphene. As electromagnetic interference due to multiple reflections only accounts for as much as 30% in the $I_{2D}/I_G$ ratio variation, there is another origin for this ratio variation. These calculations are detailed as supplementary information. $\Gamma_{2D}$ was found in the range 35 - 58 cm$^{-1}$, centered at about 40 (50) cm$^{-1}$ for top (side) pushed graphene, in the same range to what was observed for misoriented stacked MLG in.[28] Data points are displayed in a $\Gamma_{2D}$ vs $\Gamma_G$ plot in Figure 2b. The side pushed graphene data points are gathered around $\Gamma_G$ = 25 cm$^{-1}$ (in the range 18 – 26 cm$^{-1}$) and the top pushed graphene data points are gathered around $\Gamma_G$ = 20 cm$^{-1}$ (in the range 18 – 22 cm$^{-1}$), presenting a shift in $\Gamma_G$ of 4 to 12 cm$^{-1}$ from the pristine graphene data point. They are compared to graphitization trajectories obtained with carbonaceous materials such as anthracene, pitch and saccharose cokes, fibers and pyrocarbons.[55] Lespade et al. distinguished two branches: the 2-d branch corresponding to in-plane ordering and the 3-d branch corresponding to stacking ordering. The 2D band was found Lorentzian for the 2-d branch (with interplane distances in the range 0.34 - 0.35 nm) and composed of several bands for 3-d branch (with interplane distances in the range 0.33 - 0.34 nm). These results are in agreement with[40] which also proved that the global shape of the 2D band for graphitic materials depends on the way graphene planes are stacked. The pushed graphene data points stand at lower $\Gamma_G$ and $\Gamma_{2D}$ values than those reported in,[55] the side pushed graphene being the closest to the 3-d graphitic trajectory with $\Gamma_{2D}$ values up to 58 cm$^{-1}$. This plot shows that our MLG are far from graphitic materials and the trends observed are consistent with the previous observations that the pushed graphene can be considered as wi-MLG, with part of the side pushed graphene (that is close to the 3-d branch) probably better stacked and with a lower interplane distance.

**Refined Raman analysis: role of defects and competition between inner and outer scattering processes**

We now go deeper into the Raman analysis by taking advantage of the fact that the graphene pushed on the top of the squares is not perfectly homogeneous at the micrometric scale. The spectroscopic parameters related to structural, electronic or mechanical properties then vary smoothly from one edge to the other edge of the top pushed graphene zones. Figure 3 displays $I_D/I_G$ and $\Gamma_G$ as a function of d, the distance from left to right of 6 typical top pushed graphene zones (tip force equal to 1.5, 4.2, 8.1, 9.4, 10.3 and 11.1 µN). As both $I_D/I_G$ and $\Gamma_G$ are both sensitive to defects, we use these parameters together,[39] $I_D/I_G$ and $\Gamma_G$ are higher at the edges (where $I_D/I_G$ and $\Gamma_G$ can be as high as 0.3 and 28 cm$^{-1}$, respectively) than at the center (where $I_D/I_G$ and $\Gamma_G$ can be as low as 0.03 and 16.5 cm$^{-1}$, respectively), indicating that the top pushed MLG is less defective at the center than at the edges. Grey boxes in Figure 3 emphasize in average the zone centers and it can be observed that both $I_D/I_G$ and $\Gamma_G$ values are larger for the three smallest tip forces (Figures 3a and 3c) than for the three largest (Figures 3b and 3d). This indicates that, at the zone centers, more defects are created for the lowest forces. Considering that the zone edges can be significantly more affected by the cut and push process than the zone centers, we will focus in what follows on the zone center properties and the 1.5, 4.2, 8.1 µN zones will be referred to as high defect zones, while the 9.4, 10.3 and 11.1 µN zones will be referred to as low defect zones. The origin of these defects is not obvious. They can be zero-dimensional (0-d), such as dopant or vacancies, or one-dimensional (1-d), such as crystallites borders or dislocations. These two types of defects can be associated to two lengths $L_D$ and $L_a$, respectively. To retrieve $L_D$ and $L_a$ values, we use the Figure 2b of [39] and plot $A_D/A_G \times E_L^4$, ($E_L$ being the laser energy used) as a function of $\Gamma_G$ (inset of Figure 3d). Note that A refers to the band integrated area while I refers to the band heights, and that here $A_D/A_G$ approximately equals $2.2 \times I_D/I_G$. Data are gathered in the low part both of the $\Gamma_G$ values ($16 < \Gamma_G$ (cm$^{-1}$) $< 23$) and of the $A_D/A_G \times E_L^4$ values ($1 < A_D/A_G \times E_L^4$ (10 eV$^4$) $< 15$). According to the analysis of [39], in this region, $\Gamma_G$ is mainly sensitive to $L_a$ while $A_D/A_G \times E_L^4$ is mainly sensitive to $L_D$. Here we can estimate that $40 < L_a$ (nm) $< 60$ and $L_D > 30$ nm, with most of the data points corresponding to $L_D \geq 500$ nm. This analysis shows that the tip cut and push process produces defects, most probably of the two types 0-d and 1-d, but that the main effect concerns the limitation of the coherence length $L_a$ to about 50 nm. This coherence length is a few times lower than the MLG width, the latter providing a lower limit for the former, and this suggests that the source of 1-d defects is not restricted to the zone edges.

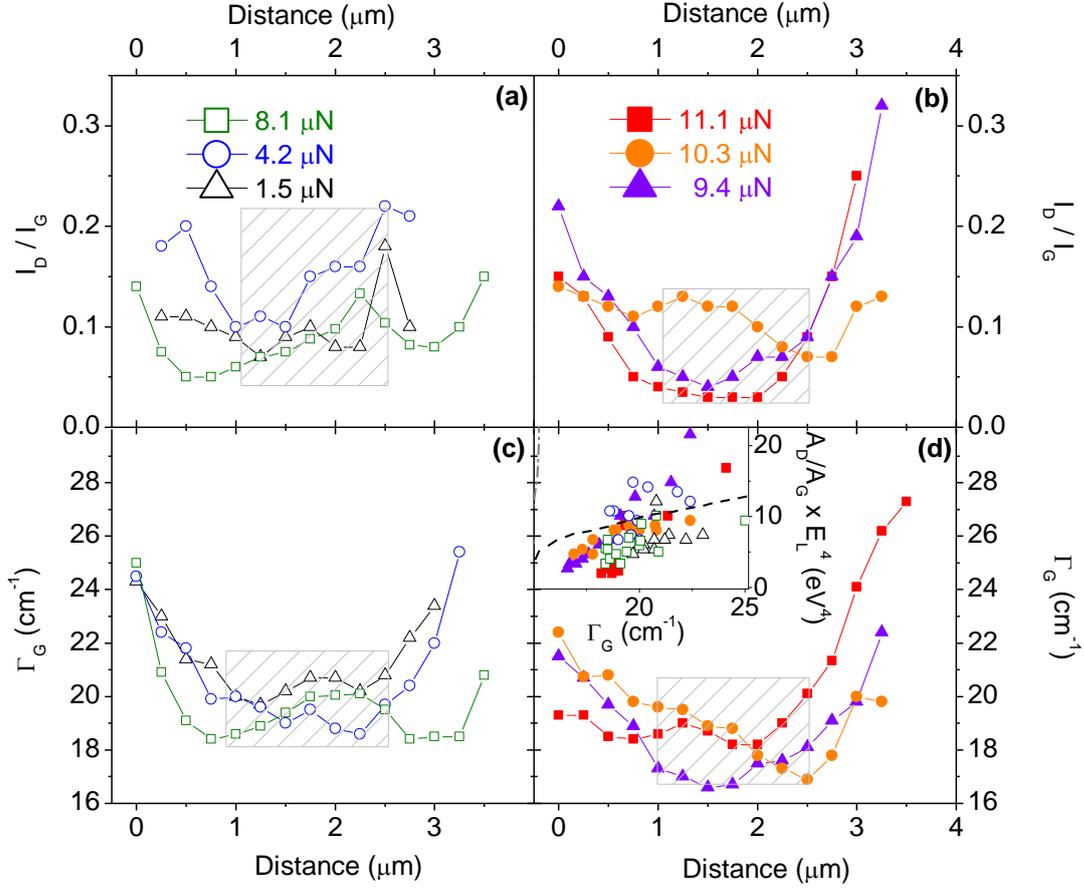

**Figure 3.** Raman profiles of top pushed graphene zones. (a) $I_D/I_G$ profiles for tip forces of 1.5, 4.2 and 8.1 μN. (b) $I_D/I_G$ profile for tip forces of 9.4, 10.3 and 11.1 μN. (c, d) same as (a) and (b) for $\Gamma_G$ profiles. Grey boxes emphasize the zone centers. The inset in Figure 3d represents $A_D/A_G \times E_L^4$ as a function of $\Gamma_G$. The dotted line is from,[39] with $L_D = 500$ nm (see text).

We now consider the analysis of the inner and outer scattering processes that compete in the shape of the 2D band by fitting it with a single or a bimodal Lorentzian (Figure 4). Spectra 1-2 are from pristine graphene and spectra 3-4, 5-6 and 7-8 are from the edge of the 8.1 μN-pushed graphene, the edge and the center of the 11.1 μN-pushed graphene, respectively. Spectra 3-4 are typical of the edges of the pushed graphene zones created with tip forces lower than ≈ 9 μN (high defect zones), whereas spectra 5-6 (7-8) are typical of the edges (centers) of the pushed graphene zones created with tip force higher than 9 μN (low defect zones). In all the cases, the 2D band residue is lower for the bimodal than for the single Lorentzian fit, which confirms the quality of the individual graphene layers and the weakness of the interlayer interaction, as discussed above.

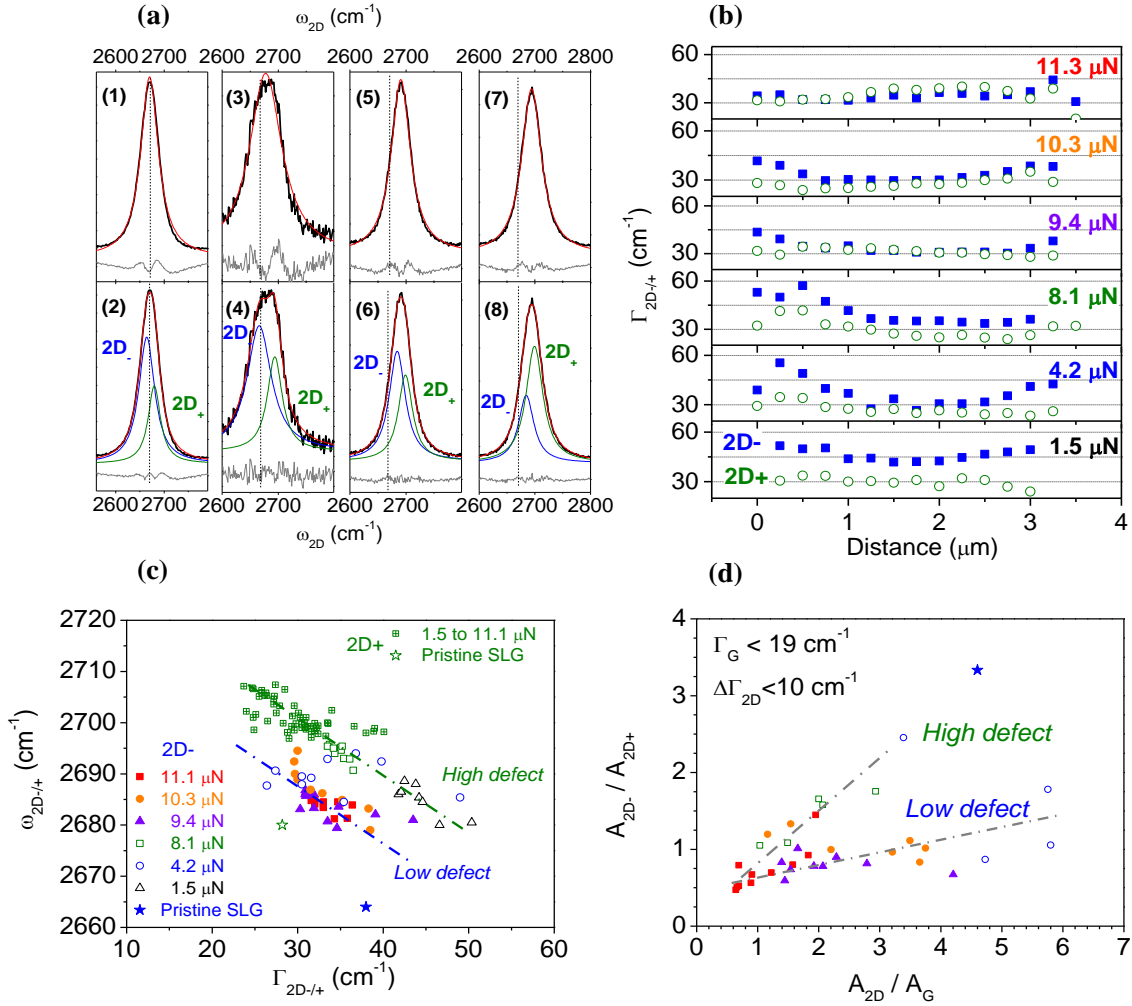

**Figure 4.** 2D$_-$ and 2D$_+$ sub bands of graphene pushed on top. (a) Fits with one (top) or two (bottom) Lorentzians for (1-2) pristine graphene, (3-4) the edge of the 8.1 μN zone, (5-6) the edge and (7-8) the center of the 11 μN zone. (b) 2D$_-$ (full square) and 2D$_+$ (empty circle) width profile for tip forces of 1.5, 4.2, 8.1, 9.4, 10.3 and 11.1 μN. (c) $\omega_{2D+/-}$ vs $\Gamma_{2D+/-}$ plot for the same tip forces, restricted to data at the zone centers (grey areas in Figure 3). (d) $A_{2D}/A_G$ vs $A_{2D-}/A_{2D+}$ plot with data from Figure 4c restricted to the conditions $\Delta\Gamma_{2D} < 10$ cm$^{-1}$ and $\Gamma_G < 19$ cm$^{-1}$. Dotted lines are guides for the eyes. Pristine SLG lies accidentally in the high defect line, as justified in the text.

A 2D. band ($\Gamma_{2D-} \approx$ 50-55 cm$^{-1}$) significantly broader than the 2D$_+$ band ($\Gamma_{2D-} \approx$30-40 cm$^{-1}$) is found for spectra 3-4 whereas 2D. and 2D$_+$ band widths are similar ($\Gamma_{2D-}$ and $\Gamma_{2D+} \approx$ 30-35 cm$^{-1}$) for spectra 7-8. The evolution of both $\Gamma_{2D-}$ and $\Gamma_{2D+}$ along the profile of each pushed graphene zone is displayed in Figure 4b. Except for the highest force (11.3 μN), the 2D. band width values are larger and more spread than for those of the 2D$_+$ band. For applied forces higher than 9 μN, $\Gamma_{2D+}$ tends to equal $\Gamma_{2D-}$ at the center of the profile. For lower forces, equality is not reached but the difference between the two widths is minimum close to the center. These trends suggest that the 2D. and 2D$_+$ widths are sensitive to defects (the lesser the number of defects, the closer the widths), the 2D. bandwidth being more sensitive than the 2D$_+$ band. Figure 4c display $\omega_{2D-}$ and $\omega_{2D+}$ as a function of $\Gamma_{2D-}$ and $\Gamma_{2D+}$ for the sampling points located at the zone centers (grey boxes in Figure 3). Remarkably, all the 2D$_+$ data points are gathered in the $\omega_{2D+}$ range from 2696 to 2708 cm$^{-1}$ and $\Gamma_{2D+}$ range from 24 to 40 cm$^{-1}$ while the 2D. data points are separated around two branches having the same slope. The upper branch contains 2D. data points from the high defect zones (tip forces below 9 μN) while the lower branch contains most of the 2D. data points from the low defect zones (tip forces above 9 μN, some points from the zone at 10.3 μN lying in between the two branches). These results confirm that the 2D band characteristics are sensitive to structural defects and that the 2D. band is more sensitive than the 2D$_+$ band. Branch 2 is downshifted horizontally by $\Delta\Gamma \approx$ 12 cm$^{-1}$ and vertically by $\Delta\omega \approx$ 15 cm$^{-1}$. It has been shown in [31] that $\Gamma_{2D-}$ and $\Gamma_{2D+}$ depend on $\gamma_{i,o}$, and $v_{Fi,o}$ where $\gamma_i$ ($\gamma_o$) is the broadening of the electron state involved in the inner (outer) scattering processes, composed of several contributions (electron-electron, electron-phonon and defect) and $v_{Fi}$ ($v_{Fo}$) is the Fermi velocity of the corresponding electronic band. Our results indicate that $v_F$, or $\gamma$, or most probably both of them depend on the graphene structural defects. The reduction of the Fermi velocity, previously discussed with Figure 2a, could explain here the vertical upshift (increase of the 2D. band position) from the low to the high defect branch. Moreover, increasing the amount of defects is expected to increase $\gamma$ and this could explain the existence of the two branches, depending on the defect amount. Moreover, a significant G band intensity enhancement is observed when simultaneously $\Gamma_{2D-}$ is close to $\Gamma_{2D+}$ and $\Gamma_G$ is not too large (< 19 cm$^{-1}$). Pure graphene corresponds to a ratio $A_{2D}/A_G$ higher than 6 and the G band enhancement which correspond to low $A_{2D}/A_G$ values has been attributed to a change of the quantum interferences involved in the G band intensity calculation appearing when doping[31, 38, 56-58]. Here, a similar process likely exists as the two conditions diminish the value of γ which appears in the denominator of the G band Raman cross section and thus increases the intensity,[57] but structural defects and not doping are at the origin of the interference changes. Figure 4d plots the intensity ratio $A_{2D-}/A_{2D+}$ as a function of $A_{2D}/A_G$ obtained for samples obeying to the two conditions: $\Delta\Gamma_{2D} < 10$ cm$^{-1}$ and $\Gamma_G < 19$ cm$^{-1}$. $A_{2D}/A_G$ values range from 0.5 to 6, emphasizing $A_G$ changes, $A_{2D}$ being roughly constant. Two straight lines are obtained, one drawn mainly by high defect zones, the other one drawn mainly by low defect zones. Pristine graphene lies in the continuity of on one of the lines, but it is known that the $A_{2D}/A_G$ for pristine graphene can vary by up to 600 % when ripples are suppressed,[59] suggesting that the alignment with the high defect line and pristine graphene is accidental. This plot strongly suggests first that there is a correlation between inner and outer process changes and the G band quantum interference changes, and second, that these changes are induced by structural defects. The decrease of $A_{2D-}/A_{2D+}$ when $A_G$ increases indicates that, while inner processes are dominant for SLG (2D. dominant), structural defects make outer processes to prevail. These results echo the recent work of[60] who studied graphene on Cu and Cu$_2$O and observed a 2D band blue shift that cannot be attributed to neither strain nor doping effects and that has been related to outer processes, highlighting the need of new theoretical calculations to understand experiments.

**Conclusion**

We report on the formation of undoped weakly interacting multilayer graphene (wi-MLG) by using nanomechanical folding induced by AFM tip on CVD transferred graphene. Vertical forces in the μN range were applied on 4 micrometer square zones that induced a cut and push process leading to the formation of wi-MLG. We obtained a large set of data to characterize this wi-MLG using Raman microscopy that allowed us to play with the amount of structural defects. The spectroscopic analysis indicates that this wi-MLG behaves neither as graphite nor as well stacked graphene, but intermediately, as graphene sheets being in weak interaction, the in plane graphene quality being maintained after folding, with large aromatic domains limited by the edges of the folded sheets. The 2D band was found composed of the two sub bands 2D$_-$ and 2D$_+$ originating in the so called inner and outer processes involved close to the Dirac cones, and blue shifted up to 20 cm$^{-1}$, suggesting a Fermi velocity reduction. Although the 2D band exists independently of structural defects involving two phonons in the double resonance mechanism, its shape was found here to depend on them, the 2D$_-$ band being much more dependent than the 2D$_+$ band. For the samples containing the lowest amount of structural defects, the 2D$_+$ band intensity was found larger than that of the 2D- band, and a correlation with changes in the G band intensity was observed. These results suggest that inner and outer processes and G band quantum interferences are related, and that they are most probably both related to the amount or the nature of structural defects. This point needs to be confirmed by additional and complementary experiments and future theoretical efforts. This work could complete the recent field of graphene origami/kirigami consisting in controlling the 3-d shape to manipulate electronic and material properties and help in better characterizing 3-d graphene involved in many applications.

**List of abbreviations:**

$\omega_x$ (expressed in cm$^{-1}$): Band Position of the band labelled X (X could be G, D, 2D, D', etc.).

$\Gamma_x$ (expressed in cm$^{-1}$): Full Width at Half Maximum of the band labelled X

$I_x$ (expressed in arbitrary units related to the number of counts on the detector): height of the band labelled X.

$A_x$ (expressed in arbitrary units related to the number of counts on the detector): integrated area of the band labelled X.

$E_L$: energy (expressed in eV) of the laser used for Raman analyses.

**SLG**: single layer graphene

**BLG**: bi-layer graphene

**MLG**: multi-layer graphene

**AFM**: atomic force microscope

**Acknowledgments**: Funding by the Spanish Ministerio de Economía y Competitividad under Project MAT2015-65356-C3-1-R is acknowledged. French PACA Region is acknowledged.

**Supporting Information Available:** Details about calculations of multiple internal reflections in multilayered films and graphene deposition details are given in a separated file.